\documentstyle[dina4,11pt]{article}

\begin{document}

\pagestyle{plain}

\newcommand{\be}{\begin{equation}}
\newcommand{\ee}{\end{equation}}
\newcommand{\bear}{\begin{eqnarray}}
\newcommand{\ear}{\end{eqnarray}}
\newcommand{\no}{\noindent}

\renewcommand{\theequation}{\arabic{section}.\arabic{equation}}
\renewcommand{\arraystretch}{2.5}
\newcommand{\GeV}{\mbox{GeV}}
\newcommand{\cL}{\cal L}
\newcommand{\D}{\cal D}
\newcommand{\Dhalf}{{D\over 2}}
\newcommand{\Det}{\rm Det}
\newcommand{\PP}{\cal P}
\newcommand{\G}{{\cal G}}
\def\R{1\!\!{\rm R}}
\def\Eins{\mathord{1\hskip -1.5pt
\vrule width .5pt height 7.75pt depth -.2pt \hskip -1.2pt
\vrule width 2.5pt height .3pt depth -.05pt \hskip 1.5pt}}
\newcommand{\symb}{\mbox{symb}}
\renewcommand{\arraystretch}{2.5}
\newcommand{\slD}{\raise.15ex\hbox{$/$}\kern-.57em\hbox{$D$}}
\newcommand{\slG}{{{\dot G}\!\!\!\! \raise.15ex\hbox {/}}}
\newcommand{\Gd}{{\dot G}}
\newcommand{\Gund}{{\underline{\dot G}}}
\def\non{\nonumber}
\def\beqn*{\begin{eqnarray*}}
\def\eqn*{\end{eqnarray*}}
\def\sy{\scriptscriptstyle}
\def\footstrut{\baselineskip 12pt}
\def\square{\kern1pt\vbox{\hrule height 1.2pt\hbox{\vrule width 1.2pt
   \hskip 3pt\vbox{\vskip 6pt}\hskip 3pt\vrule width 0.6pt}
   \hrule height 0.6pt}\kern1pt}
\def\dint#1{\int\!\!\!\!\!\int\limits_{\!\!#1}}
\def\bra#1{\langle #1 |}
\def\ket#1{| #1 \rangle}
\def\vev#1{\langle #1 \rangle}
\def\dps{\displaystyle}
\def\sy{\scriptscriptstyle}
\def\half{{1\over 2}}
\def\third{{1\over3}}
\def\fourth{{1\over4}}
\def\fifth{{1\over5}}
\def\sixth{{1\over6}}
\def\seventh{{1\over7}}
\def\eigth{{1\over8}}
\def\ninth{{1\over9}}
\def\tenth{{1\over10}}
\def\pa{\partial}
\def\ddtau{{d\over d\tau}}
\def\ge{\hbox{\textfont1=\tame $\gamma_1$}}
\def\gz{\hbox{\textfont1=\tame $\gamma_2$}}
\def\gd{\hbox{\textfont1=\tame $\gamma_3$}}
\def\go{\hbox{\textfont1=\tamt $\gamma_1$}}
\def\gt{\hbox{\textfont1=\tamt $\gamma_2$}}
\def\gth{\hbox{\textfont1=\tamt $\gamma_3$}} 
\def\gf{\hbox{$\gamma_5\;$}}
\def\ie{\hbox{$\textstyle{\int_1}$}}
\def\iz{\hbox{$\textstyle{\int_2}$}}
\def\id{\hbox{$\textstyle{\int_3}$}}
\def\ldop{\hbox{$\lbrace\mskip -4.5mu\mid$}}
\def\rdop{\hbox{$\mid\mskip -4.3mu\rbrace$}}
\def\e{\mbox{e}}
\def\g{\mbox{g}}
\def\pa{\partial}
\def\kinb{{1\over 4}\dot x^2}
\def\kinf{{1\over 2}\psi\dot\psi}
\def\expk{{\rm exp}\biggl[\,\sum_{i<j=1}^4 G_{Bij}k_i\cdot k_j\biggr]}
\def\expp{{\rm exp}\biggl[\,\sum_{i<j=1}^4 G_{Bij}p_i\cdot p_j\biggr]}
\def\expshort{{\e}^{\half G_{Bij}k_i\cdot k_j}}
\def\expabb{{\e}^{(\cdot )}}
\def\Dab{{(x_a-x_b)}}
\def\Dsq{{({(x_a-x_b)}^2)}}
\def\lag{( -\partial^2 + V)}
\def\Tintm{{\dps\int_{0}^{\infty}}{dT\over T}e^{-m^2T}}
\def\Tint{{\dps\int_{0}^{\infty}}{dT\over T}}
\def\pint{{\dps\int}{dp_i\over {(2\pi)}^d}}
\def\Dx{\dps\int{\cal D}x}
\def\Dy{\dps\int{\cal D}y}
\def\Dpsi{\dps\int{\cal D}\psi}
\def\Tr{{\rm Tr}\,}
\def\tr{{\rm tr}\,}
\def\freeexp{{\rm e}^{-\int_0^Td\tau {1\over 4}\dot x^2}}
\def\arraystretch{2.5}
\def\Ge{\mbox{GeV}}
\def\dA{\partial^2}
\def\DA{\sqsubset\!\!\!\!\sqsupset}
%
\def\bbbr{{\rm I\!R}}
\def\bbbone{{\mathchoice {\rm 1\mskip-4mu l} {\rm 1\mskip-4mu l}
{\rm 1\mskip-4.5mu l} {\rm 1\mskip-5mu l}}}
\def\bbbz{{\mathchoice {\hbox{$\sf\textstyle Z\kern-0.4em Z$}}
{\hbox{$\sf\textstyle Z\kern-0.4em Z$}}
{\hbox{$\sf\scriptstyle Z\kern-0.3em Z$}}
{\hbox{$\sf\scriptscriptstyle Z\kern-0.2em Z$}}}}
%

\pagestyle{empty}

\renewcommand{\thefootnote}{\fnsymbol{footnote}}

\hskip 10cm {\sl ANL-HEP-PR-97-83}
\vskip-.1pt
\hskip 10cm {\sl HUB-EP-97/35}

\vskip 1.8cm
\begin{center}
{\bf\Large The Structure of the Bern -- Kosower Integrand}\medskip\\
{\bf\Large for the N -- Gluon Amplitude}
\vskip1cm

\vskip 1.3cm
{\large Christian Schubert
        \footnote{e-mail address schubert@qft2.physik.hu-berlin.de} 
        \footnote{Supported by Deutsche Forschungsgemeinschaft} }
        \footnote{Work supported in part by the U.S. Department
                  of Energy, High Energy Physics Division, under
                  contract W-31-109-ENG-38.}

\vskip1.3cm 

{\it High Energy Physics Division,
Argonne National Laboratory,\\
Argonne, IL 60439-4815, USA}

\vskip.1cm
and
\vskip.1cm

{\it Institut f\"ur Physik,
Humboldt Universit\"at zu Berlin,\\
Invalidenstr. 110, D-10115 Berlin, Germany}

\vskip2.5cm

{\large\bf Abstract}

\end{center}

\begin{quotation}

\noindent
An ambiguity inherent in the partial integration procedure
leading to the Bern-Kosower rules is fixed in a way which
preserves the complete permutation symmetry in the
scattering states. This leads to a canonical version
of the Bern-Kosower representation for the one-loop
$N$ -- photon/gluon amplitudes, and to a natural
decomposition of those amplitudes into permutation
symmetric gauge invariant partial amplitudes.
This decomposition exhibits a simple recursive
structure.

\end{quotation}

\clearpage

\renewcommand{\thefootnote}{\protect\arabic{footnote}}
\pagestyle{plain}
\renewcommand{\theequation}{\arabic{equation}}
\setcounter{equation}{0}
\setcounter{page}{1}
\setcounter{footnote}{0}

In recent years it has been found
that string theory can serve as a guiding
principle for the derivation of useful and non-trivial
rearrangements in standard perturbative quantum field theory.
While such string-related techniques 
have been 
applied to a large
variety of field theory problems
~\cite{berkos:prl166,berkos:npb362,berkos:npb379,bediko5glu,bedush,dunnor,strassler1,strassler2,ss1,ss23,fss/fhss,rolsat,dashsu,mnss,dhogag,sato,rss,dlmmr}
the primary example is still the case of the 
one -- loop $N$ -- photon or -- gluon amplitudes.  
A recipe for the construction of this amplitude is given
by the ``Bern -- Kosower
Rules'', which originally were derived by an analysis of the
infinite string tension limit of the corresponding amplitude
in an appropriate string model 
~\cite{berkos:prl166,berkos:npb362,berkos:npb379}
(see ~\cite{berntasi} for a review). 
A simpler derivation of the same rules was later given
by Strassler 
in the so-called world line path integral formalism
~\cite{strassler1,strassler2,ss1,ss23,dashsu,mnss,dhogag,sato,rss,polbook} (see ~\cite{zako} for an introductory exposition).
In this approach one represents one -- loop effective
actions in standard quantum field theory in terms
of certain first -- quantized particle path integrals, and
evaluates those in a way analogous to the calculation of the
Polyakov path integral in string theory.
The path integral relevant for the $N$ -- photon/gluon amplitude
is the following ~\cite{feynman:pr80,strassler1}

\begin{equation}
\Gamma\lbrack A\rbrack = 
{\rm tr}
\int_0^{\infty}
{dT\over T} \, {\rm e}^{-m^2T}
\int {\cal D}x\, 
{\cal P}
{\rm exp} \left[ 
- \int_0^T \!\!\! d\tau \left( {1\over 4}{\dot x}^2 
+ igA_{\mu}\dot x^{\mu} 
\right) \right]
\label{scalarpi}
\end{equation}
\no
This formula expresses the one-loop effective action
induced by a (complex) scalar loop 
with mass $m$ for a Yang-Mills
background field in terms of a quantum mechanical
path integral. At fixed Schwinger proper-time
$T$, the path integral is to be performed
over the space of trajectories obeying
$x(T)=x(0)$. 
${\rm tr}$ denotes the global colour trace,
and $\cal P$ the path-ordering of the exponential
(those can be omitted in the
abelian case). We use Euclidean conventions.
Similar path integral representations exist for the
fermion loop ~\cite{fradkin,borcas,fragit} 
and gluon loop ~\cite{strassler1,rss}
contributions to this amplitude. 

The $N$ -- point amplitude can be extracted
from this path integral by
expanding the interaction term to $N$ - th order,
and then specializing to a background consisting
of plane waves carrying definite polarizations 
$\varepsilon_i$
and
gauge algebra generators $T^{a_i}$. 
Introducing the string theoretic photon (gluon) vertex operator 

\be
V_i= (T^{a_i})
\int_0^T d\tau_i \,\varepsilon_i\cdot\dot x(\tau_i)
{\e}^{ik_i\cdot x(\tau_i)}
\label{gluonvertop}
\ee
\no
the result can be written as (for the gluon case)

\bear
\Gamma^{a_1\ldots a_N}
[k_1,\varepsilon_1;\ldots;k_N,\varepsilon_N]
&=&{(-ig)}^N
\tr\Tint\e^{-m^2T}
\int{\cal D}x(\tau)
V_1V_2\cdots V_N
\nonumber\\
&&\times
\delta({\tau_N\over T})
\prod_{i=1}^{N-1}
\theta(\tau_i-\tau_{i+1})
{\rm exp} \left[ 
- \int_0^T \!\!\! d\tau {1\over 4}{\dot x}^2 
\right]
\non\\
\label{Ngluonwickscal}
\ear
\no
Here the zero on the loop has been fixed
to be at the location of the $N$ - th vertex operator.
The functions $\theta(\tau_i-\tau_{i+1})$ 
implement the path ordering = colour ordering
which one has in the non-Abelian case.
This path integral is Gaussian, 
so that its evaluation
can be done simply by ``completing the square''.
To get an invertible kinetic term,
first one extracts the
zero mode 
$x_0\equiv {1\over T}\int_0^T\,d\tau x(\tau)$
from the path integral. The
integral over $x_0$
is separated off,
and just
produces the usual energy-momentum conservation factor.
The remaining path integral is then performed using the
worldline Green's function

\be
G_B(\tau_1,\tau_2)=\mid \tau_1-\tau_2\mid 
-{{(\tau_1-\tau_2)}^2\over T} 
\label{defG}
\ee
\no
Rewriting

\be
\varepsilon_i\cdot
\dot x_i\e^{ik_i\cdot x_i}
=
\e^{\varepsilon_i\cdot\dot x_i
+ik_i\cdot x_i}
\mid_{{\rm lin}(\varepsilon_i)}
\label{formexp}
\ee
\no
one arrives at the following
master formula
for the scalar loop contribution to
the
one-loop
$N$ - gluon amplitude,

\begin{eqnarray}
\Gamma^{a_1\ldots a_N}
[k_1,\varepsilon_1;\ldots;k_N,\varepsilon_N]
&=&
{(-ig)}^N
{\rm tr}
(T^{a_1}\cdots T^{a_N})
{(2\pi )}^D\delta (\sum k_i)
\nonumber\\
&&
\!\!\!\!\!\!\!\!\!\!\!\!\!\!\!\!\!\!
\!\!\!\!\!\!\!\!\!\!\!\!\!\!\!\!\!\!
\times
{\dps\int_{0}^{\infty}}{dT\over T}
{[4\pi T]}^{-{D\over 2}}
e^{-m^2T}
\prod_{i=1}^N \int_0^T 
d\tau_i\,
\delta({\tau_N\over T})
\prod_{i=1}^{N-1}
\theta(\tau_i-\tau_{i+1})
\nonumber\\
&&
\!\!\!\!\!\!\!\!\!\!\!\!\!\!\!\!\!\!
\!\!\!\!\!\!\!\!\!\!\!\!\!\!\!\!\!\!
\times
\exp\biggl\lbrace\sum_{i,j=1}^N 
\Bigl\lbrack 
\half G_{Bij} k_i\cdot k_j
-i\dot G_{Bij}\varepsilon_i\cdot k_j
+\half\ddot G_{Bij}\varepsilon_i\cdot\varepsilon_j
\Bigr\rbrack\biggr\rbrace
\mid_{\rm multi-linear}
\nonumber\\
\label{scalarqedmaster}
\end{eqnarray}
\no
Here it is understood that only the terms linear
in all the $\varepsilon_1,\ldots,\varepsilon_N$
have to be taken. 
Besides the Green's function $G_B$ also its first and
second derivatives appear,

\begin{eqnarray}
\dot G_B(\tau_1,\tau_2) &=& {\rm sign}(\tau_1 - \tau_2)
- 2 {{(\tau_1 - \tau_2)}\over T}\nonumber\\
\ddot G_B(\tau_1,\tau_2)
&=& 2 {\delta}(\tau_1 - \tau_2)
- {2\over T}\quad \nonumber\\
\label{GdGdd}
\end{eqnarray}
\noindent
Dots generally denote a
derivative acting on the first variable,
$\dot G_B(\tau_1,\tau_2) \equiv {\partial\over
{\partial\tau_1}}G_B(\tau_1,\tau_2)$, 
and we abbreviate
$G_{Bij}\equiv G_B(\tau_i,\tau_j)$ etc.

\noindent
Writing out the exponential in eq.(\ref{scalarqedmaster})
one obtains an integrand

\be
\exp\biggl\lbrace
\cdot
\biggr\rbrace
\mid_{\rm multi-linear}
\quad={(-i)}^N
P_N(\dot G_{Bij},\ddot G_{Bij})
\exp\biggl[\half
\sum_{i,j=1}^N G_{Bij}k_i\cdot k_j
\biggr]
\label{defPN}
\ee\no
with a certain polynomial $P_N$ depending on the various 
$\dot G_{Bij},\ddot G_{Bij}$ and on the kinematic invariants.
The resulting parameter integrals are directly
related to the ones arising in a standard 
Feynman parameter
calculation of this amplitude ~\cite{berdun,strassler1,morgan}.
The exponential factor in particular will, after
performance of the global $T$ -- integration, turn
into the standard one-loop 
$N$ - point Feynman denominator polynomial.
To arrive at the Bern - Kosower rules, one now has to 
remove all second derivatives $\ddot G_{Bij}$ appearing
in $P_N$
by suitable partial integrations in the variables
$\tau_i$,

\be
P_N(\dot G_{Bij},\ddot G_{Bij})
\e^{\half\sum G_{Bij}k_i\cdot k_j}
\quad
{\stackrel{\sy{\rm part. int.}}{\longrightarrow}}
\quad
Q_N(\dot G_{Bij})
\e^{\half\sum G_{Bij}k_i\cdot k_j}
\label{partint}
\ee\no
That this is possible for any $N$ was proven in appendix B
of ~\cite{berkos:npb362}. The new integrand is written
in terms of the $G_{Bij}$ and $\dot G_{Bij}$ alone, and
serves as the input for the Bern - Kosower
rules. Those allow one to classify the various 
contributions to the $N$ -- photon/gluon amplitude
in terms of $\phi^3$ -- diagrams, and moreover
lead to simple relations between the integrands 
for the scalar, spinor and gluon loop cases.
A complete formulation of the rules is lengthy,
and we refer the reader to
~\cite{berkos:npb379,berntasi}.
Let us just remark that, up to global factors correcting
for the differences in degrees of freedom and
statistics, the integrand for the spinor
loop case can be obtained from the one for
the scalar loop simply
by replacing every closed cycle of $\dot G_B$'s
appearing in $Q_N$ by its
``supersymmetrization'',

\vspace{-10pt}
\begin{equation}
\dot G_{Bi_1i_2} 
\dot G_{Bi_2i_3} 
\cdots
\dot G_{Bi_ni_1}
\rightarrow 
\dot G_{Bi_1i_2} 
\dot G_{Bi_2i_3} 
\cdots
\dot G_{Bi_ni_1}
-
G_{Fi_1i_2}
G_{Fi_2i_3}
\cdots
G_{Fi_ni_1}
\nonumber\\
\label{subrulechap5}
\end{equation}
where $G_{F12}={\rm sign}(\tau_1-\tau_2)$
denotes the fermionic worldline Green's
function. Note that an expression is considered a cycle
already if it can be put into cycle form
using the antisymmetry of $\dot G_B$ (e.g.
$\dot G_{B12}\dot G_{B12}=-\dot G_{B12}\dot G_{B21}$).
A similar ``cycle replacement rule''
holds for the gluon loop case.

Our objective in this letter is a further investigation of
the partial integration procedure, and of the
structure of the polynomial $Q_N$. 
We begin with the two-point
amplitude. For $N=2$ eq.(\ref{defPN}) yields

\be
P_2= \dot G_{B12}\varepsilon_1\cdot k_2
\dot G_{B21}\varepsilon_2\cdot k_1
-\ddot G_{B12}\varepsilon_1\cdot\varepsilon_2
\label{P2}
\ee\no
After a partial integration performed
on the second term in
$\tau_1$ or $\tau_2$ this turns into

\be
Q_2= 
\Bigl\lbrack
\varepsilon_1\cdot k_2\varepsilon_2\cdot k_1
-\varepsilon_1\cdot\varepsilon_2
k_1\cdot k_2
\Bigr\rbrack
\dot G_{B12}\dot G_{B21}
\label{Q2}
\ee\no
We note 
the following two effects
of this partial integration:

\begin{enumerate}

\item
The new Feynman numerator polynomial is
a function of $\dot G_{B12}$, and homogeneous
in the external momenta $k_i$.

\item
A transversal projector has appeared,
making
the gauge invariance manifest
at the integrand level.

\end{enumerate}
%
In the three - point case one finds

\be
P_3= 
\dot G_{B1i}\varepsilon_1\cdot k_i
\dot G_{B2j}\varepsilon_2\cdot k_j
\dot G_{B3k}\varepsilon_3\cdot k_k
- 
\Bigl[
\ddot G_{B12}\varepsilon_1\cdot\varepsilon_2
\dot G_{B3i}\varepsilon_3\cdot k_i
+ 2\, {\rm permuted \,\, terms}
\Bigr] 
\non
\label{P3}
\ee\no
Here and in the following the dummy indices
$i,j,k$
should be summed over
from $1$ to $N$, and one has
$\dot G_{Bii}=0$ by antisymmetry.
Removing all the $\ddot G_{Bij}$'s
by partial integrations one finds

\bear
Q_3&=&
\dot G_{B1i}\varepsilon_1\cdot k_i
\dot G_{B2j}\varepsilon_2\cdot k_j
\dot G_{B3k}\varepsilon_3\cdot k_k
\non\\
&&
+\half
\biggl\lbrace
\dot G_{B12}
\varepsilon_1\cdot\varepsilon_2
\biggl\lbrack
\dot G_{B3i}
\varepsilon_3\cdot k_i
\bigl(
\dot G_{B1j}k_1\cdot k_j
-\dot G_{B2j}k_2\cdot k_j
\bigr)
\non\\
&&
+
\bigl(
\dot G_{B31}
\varepsilon_3\cdot k_1
-\dot G_{B32}
\varepsilon_3\cdot k_2
\bigr)
\dot G_{B3j}
k_3\cdot k_j
\biggr\rbrack
+ 2\,\, {\rm perm.}
\biggr\rbrace
\non\\
&=& Q_3^3 + Q_3^2 \non\\
\label{Q3}
\ear\no
where

\bear
Q_3^3&=&
\dot G_{B12}\dot G_{B23}\dot G_{B31}
Z_3(123)\non\\
Q_3^2&=&
\dot G_{B12}\dot G_{B21}
Z_2(12)
\dot G_{B3i}\varepsilon_3\cdot k_i
+ 2 \,\, {\rm perm.}
\non\\
\label{Q3components}
\ear\no
We have now introduced the notation

\bear
Z_2(ij)&\equiv&
\varepsilon_i\cdot k_j
\varepsilon_j\cdot k_i
-\varepsilon_i\cdot\varepsilon_j
k_i\cdot k_j
\non\\
Z_n(i_1i_2\ldots i_n)&\equiv&
{\rm tr}
\prod_{j=1}^n 
\Bigl[
k_{i_j}\otimes \varepsilon_{i_j}
- \varepsilon_{i_j}\otimes k_{i_j}
\Bigr]
\quad (n\geq 3)
\label{defZn}
\ear\no
for the cyclically invariant Lorentz tensors which
appear in the result.
$Z_n$ corresponds to a ${\rm tr}[F^n]$ in the 
(abelian) effective action,
and after the partial integration procedure appears multiplied by
a factor of $\dot G_{Bi_1i_2}\dot G_{Bi_2i_3}\cdots
\dot G_{Bi_ni_1}$, independently of the algorithm used
~\cite{strassler2}. 
The ``$\tau$ -- cycles''
appearing in the Bern-Kosower
substitution rules are thus 
associated to
the ``Lorentz cycles''.

In the abelian case the three photon
amplitude must vanish by Furry's theorem. 
To verify that this is indeed the case note that
the integrand is odd under the
transformation of variables
$\tau_i=T-\tau_i'$,
$i=1,2,3$,
since

\be
G_B(\tau_i,\tau_j)=
G_B(\tau_i',\tau_j'),
\quad
\dot G_B(\tau_i,\tau_j)=
-\dot G_B(\tau_i',\tau_j')
\label{Gmirror}
\ee\no
In the three-point case,
$Q_3$ is still unique; all possible
ways of performing the partial integrations
lead to the same result.
The same is not true any more in the
four-point case, where the result
of the partial integration procedure
turns out to depend on the
specific chain of partial integrations
chosen. 
This ambiguity was discussed in
~\cite{strassler2}, 
and the question
asked whether some particular algorithm
exists which would 
not single out any of the variables $\tau_i$,
and thus
preserve the full permutation
symmetry between the $N$ external legs.

\noindent
We will now define such an 
``impartial'' partial integration
algorithm, in the following way:

\begin{enumerate}
\item
In every step, partially integrate away {\sl all}
$\ddot G_{Bij}$'s appearing in the
term under inspection {\sl simultaneously}.
This is possible since different $\ddot G_{Bij}$'s do
not share variables to being with, and this property
is preserved by all partial integrations. New
$\ddot G_{Bij}$'s may be created.

\item
In the first step, for every $\ddot G_{Bij}$ partially
integrate both over $\tau_i$ and $\tau_j$,
and take the mean of the results.

\item
At every following step, any $\ddot G_{Bij}$
appearing must have been created in the 
previous step. Therefore either both $i$ and $j$
were partially
integrated over in the previous step, or just
one of them. If both, the rule is to again use both
variables in the actual step for partial integration,
and take the mean of the results. If only one of them
was used
in the previous step, 
then the other one should be used in the actual
step.

\end{enumerate}
\no
For example, the term
$\ddot G_{B12}\ddot G_{B34}$
appearing in $P_4$ 
in the first step transforms as follows,

\bear
\ddot G_{B12}\ddot G_{B34}&\rightarrow&
\fourth
\dot G_{B12}\dot G_{B34}
\biggl\lbrace
\Bigl[
\dot G_{B1i}k_1\cdot k_i-\dot G_{B2i}k_2\cdot k_i
\Bigr]
\Bigl[
\dot G_{B3j}k_3\cdot k_j-\dot G_{B4j}k_4\cdot k_j
\Bigr]
\non\\
&&
-\ddot G_{B13}k_1\cdot k_3
 +\ddot G_{B14}k_1\cdot k_4 +\ddot G_{B23}k_2\cdot k_3
-\ddot G_{B24}k_2\cdot k_4
\biggr\rbrace
\non\\
\label{alg1step1}
\ear\no
The terms in the second line have to be further processed.
Considering just the first one of them, since both variables
appearing in $\ddot G_{B13}$ were active in the
first step, both must also be used in the second one.
This yields

\bear
-\fourth
\dot G_{B12}\dot G_{B34}
\ddot G_{B13}
&\rightarrow&
\eigth
\dot G_{B12}\dot G_{B34}
\dot G_{B13}
\Bigl[
\dot G_{B1i}k_1\cdot k_i
-\dot G_{B3i}k_3\cdot k_i
\Bigr]
\non\\
&&
+\eigth
\dot G_{B13}
\Bigl[
\ddot G_{B12}\dot G_{B34}
-\dot G_{B12}\ddot G_{B34}
\Bigl]
\non\\
\label{alg1step2}
\ear\no
Considering again the first term in the second
line, only $\tau_1$ was active in the previous step.
Therefore only $\tau_2$ must be used now, and the
third step is the final one,

\be
\eigth
\dot G_{B13}\ddot G_{B12}\dot G_{B34}
\rightarrow
\eigth
\dot G_{B13}\dot G_{B12}\dot G_{B34}
\dot G_{B2i}k_2\cdot k_i
\label{alg2step3}
\ee\no
This prescription treats all variables on the same
footing, and therefore must lead to a permutation
symmetric result.
The nontrivial fact is that the process terminates after a finite
number of steps, and does not become cyclic (as would be the case if,
for example, one would {\it always} treat the indices
in a $\ddot G_{Bij}$ symmetrically). 
This 
is not difficult to derive from the fact that, 
for any term in $P_N$,
the indices appearing in the $\ddot G_{Bij}$'s
and the
first indices of the $\dot G_{Bij}$'s
are associated to the polarization vectors, and thus
must all take different values.

\noindent
This algorithm transforms $P_4$ into
\bear
Q_4&=&
\dot G_{B1i}\varepsilon_1\cdot k_i
\dot G_{B2j}\varepsilon_2\cdot k_j
\dot G_{B3k}\varepsilon_3\cdot k_k
\dot G_{B4l}\varepsilon_4\cdot k_l
\non\\
&&
\!\!\!\!\!\!\!\!\!
+
\Biggl\lbrace
\half\dot G_{B12}
\varepsilon_1\cdot \varepsilon_2
\biggl\lbrace
\dot G_{B3i}\varepsilon_3\cdot k_i
\dot G_{B4j}\varepsilon_4\cdot k_j
\Bigl[
\dot G_{B1k}k_1\cdot k_k
-
\dot G_{B2k}k_2\cdot k_k
\Bigr]
\non\\
&&\!\!\!
+
\Bigl[
\dot G_{B3i}\varepsilon_3\cdot k_i
\bigl(
\dot G_{B41}\varepsilon_4\cdot k_1
-
\dot  G_{B42}\varepsilon_4\cdot k_2
\bigr)
\dot G_{B4k}k_4\cdot k_k
+ \bigl( 3 \leftrightarrow 4)
\Bigr]
\non\\
&&\!\!\!
+
\Bigl[
\bigl(
\dot G_{B31}\varepsilon_3\cdot k_1
-
\dot  G_{B32}\varepsilon_3\cdot k_2
\bigr)
\dot G_{B43}\varepsilon_4\cdot k_3
\dot G_{B4k}k_4\cdot k_k
+ \bigl( 3 \leftrightarrow 4)
\Bigr]
\biggr\rbrace
+ 5\,\, {\rm permutations} 
\Biggr\rbrace
\non\\
&&\!\!\!\!\!\!\!\!
+
\Biggl\lbrace
\fourth
\dot G_{B12}\dot G_{B34}
\varepsilon_1\cdot\varepsilon_2
\varepsilon_3\cdot\varepsilon_4
\biggl\lbrace
\Bigl[
\dot G_{B1i}k_1\cdot k_i
-\dot G_{B2i}k_2\cdot k_i
\Bigr]
\Bigl[
\dot G_{B3j}k_3\cdot k_j
-\dot G_{B4j}k_4\cdot k_j
\Bigr]
\non\\
&&\!\!\!
+\half
\Bigl[
\dot G_{B13}k_1\cdot k_3
-\dot G_{B23}k_2\cdot k_3
-\dot G_{B14}k_1\cdot k_4
+\dot G_{B24}k_2\cdot k_4
\Bigr]
\non\\
&&\!\!\!
\times
\Bigl[
\dot G_{B1i}k_1\cdot k_i
+\dot G_{B2i}k_2\cdot k_i
-\dot G_{B3i}k_3\cdot k_i
-\dot G_{B4i}k_4\cdot k_i
\Bigr]
\biggr\rbrace
+ 2\,\, {\rm perm.} 
\Biggr\rbrace
\label{Q4}
\ear\no
This expression can be rewritten 
more compactly as follows,

\be
Q_4
=
Q_4^4 + Q_4^3 + Q_4^2 - Q_4^{22}
\non\\
\label{4photon}
\ee\no
where
\bear
Q_4^4 &=& 
\dot G_{B12}
\dot G_{B23}
\dot G_{B34}
\dot G_{B41}
Z_4(1234)
+ 2 \,\, {\rm permutations}
\non\\
Q_4^3 &=&
\dot G_{B12}
\dot G_{B23}
\dot G_{B31}
Z_3(123)
\dot G_{B4i}
\varepsilon_4\cdot k_i
+ 3 \,\, {\rm perm.}
\non\\
Q_4^2 &=&
\dot G_{B12}\dot G_{B21}
Z_2(12)
\biggl\lbrace
\dot G_{B3i}
\varepsilon_3\cdot k_i
\dot G_{B4j}
\varepsilon_4\cdot k_j
+\half
\dot G_{B34}
\varepsilon_3\cdot\varepsilon_4
\Bigl[
\dot G_{B3i}
k_3\cdot k_i
-
\dot G_{B4i}
k_4\cdot k_i
\Bigr]
\biggr\rbrace
\non\\
&&
+ \, 5 \,\, {\rm perm.}
\non\\
Q_4^{22} &=&
\dot G_{B12}\dot G_{B21}
Z_2(12)
\dot G_{B34}\dot G_{B43}
Z_2(34)
+ 2 \,\, {\rm perm.}
\non\\
\label{4photoncoeff}
\ear\no
This decomposition according to cycles
is not only necessary for the application
of the Bern-Kosower substitution rules, but also
natural in terms of gauge invariance.
The sixteen terms appearing in this
decomposition are individually
gauge invariant, i.e. they either
vanish or turn into total derivatives
if the replacement
$\varepsilon_i \rightarrow k_i$
is made for any of the external legs.
This is trivial for $Q_4^4, Q_4^{22}$, and
also for $Q_4^3$, since if we substitute
$k_4$ for $\varepsilon_4$ there (in the un-permuted term)
we have a total derivative at hand,

$$\partial_4
\biggl[
\dot G_{B12}
\dot G_{B23}
\dot G_{B31}
{\e}^{\half G_{Bij}k_i\cdot k_j}
\biggr]
$$
\no
The only not quite trivial case is 
a replacement of $\varepsilon_3$ 
or $\varepsilon_4$ in (the un-permuted term of)
$Q_4^2$. 
By inspection one finds that 
the replacement $\varepsilon_3\to k_3$
yields
the total derivative

\bear
\partial_3
\biggl[
\dot G_{B12}\dot G_{B21}
Z_2(12)
\dot G_{B4j}
\varepsilon_4\cdot k_j
\,{\e}^{\half G_{Bij}k_i\cdot k_j}
\biggr]
+\half (\partial_3-\partial_4)
\biggl[
\dot G_{B12}\dot G_{B21}
Z_2(12)
\dot G_{B34}k_3\cdot\varepsilon_4
\,{\e}^{(\cdot )}
\biggr]
\non\\
\label{Q4totderiv}
\ear
\no
and analogously for $\varepsilon_4$.
Note that the product of two-cycles $Q_4^{22}$
appears with a minus sign in eq.(\ref{4photon}). 
The reason is that we corrected for an over-counting
here; $Q_4^{22}$ is also contained twice
in $Q_4^2$, and 
separating it out from there will change the
``-'' to a ``+''.

Before proceeding to higher point amplitudes,
it will be prudent to
further condense the notation.
We thus abbreviate

\bear
\dot G_{ij}&\equiv & \dot G_{Bij}
\varepsilon_i\cdot k_j\non\\
\Gund_{ij}
&\equiv &
\dot G_{Bij}\varepsilon_i\cdot\varepsilon_j
\non\\
{{\slG}_{ij}}
&\equiv &
\dot G_{Bij}
k_i\cdot k_j\non\\
{\dot G}(i_1i_2\ldots i_n)
&\equiv&
\dot G_{Bi_1i_2}
\dot G_{Bi_2i_3}
\cdots
\dot G_{Bi_ni_1}
Z_n(i_1i_2\ldots i_n)\non\\
\label{defabb}
\ear\no
As was mentioned before, it
is known from previous work
~\cite{berkos:npb362,berkos:npb379,strassler2}
that a closed ``$\tau$ -- cycle''
$\dot G_{Bi_1i_2}
\dot G_{Bi_2i_3}
\cdots
\dot G_{Bi_ni_1}$
after the partial integration will
always appear multiplied by a 
complete factor of
$Z_n(i_1i_2\ldots i_n)$. 
This motivates the last one of the
abbreviations above, and also explains why the
formulation of the ``cycle substitution'' part
of the Bern-Kosower rules did not require the 
specification of a particular partial
integration
algorithm.

A given term in $Q_N$ thus will be a product
of ``complete cycles'' $\dot G(\cdot )$,
multiplied by a remainder. Following
~\cite{strassler2} we call this remainder
``tail'', or ``$m$ - tail'', where $m$ denotes
the number of indices not appearing in 
any of the cycles. For example, $Q_4^2$ is
the product of a complete $2$ - cycle and
a $2$ - tail. 
Only the tails depend on the
choice of the partial integration algorithm.
The tail generated by our specific 
symmetric algorithm
will be denoted by $T_m(i_1\ldots i_m)$.
The $1$ - tail is (unambiguously) given by
$T_1(i)=\dot G_{ij}$ ($i$ being fixed
and $j$ summed over).

With the above abbreviations, the result for $Q_5$ 
obtained by an application of the symmetric 
algorithm can be written as follows,

\be
Q_5
=
Q_5^5 + Q_5^4 + Q_5^3 + Q_5^2 
- Q_5^{32} - Q_5^{22}
\non\\
\label{5photon}
\ee\no
where
\bear
Q_5^5 &=& 
\dot G(12345)
+ 11\,\, {\rm permutations}
\non\\
Q_5^4 &=&
\dot G(1234)
\dot G_{5i}
+ 14\,\, {\rm perm.}
\non\\
Q_5^3 &=&
\dot G(123)
\biggl\lbrace
\dot G_{4i}
\dot G_{5j}
+\half
\Gund_{45}
\Bigl[
\slG_{4i}
-
\slG_{5i}
\Bigr]
\biggr\rbrace
+ 9 \,\,{\rm perm.}
\non\\
Q_5^2 &=&
\dot G(12)
\Biggl\lbrace
\dot G_{3i}\dot G_{4j}\dot G_{5k}
+\half \Gund_{34}
\biggl[
\Gd_{5k}
\Bigl[
\slG_{3i}-\slG_{4i}
\Bigr]
+\slG_{5i}
\Bigl[
\Gd_{53}-\Gd_{54}
\Bigr]
\biggr]
\non\\
&&
\phantom{\dot G(12)}
+\half \Gund_{35}
\biggl[
\Gd_{4k}
\Bigl[
\slG_{3i}-\slG_{5i}
\Bigr]
+\slG_{4i}
\Bigl[
\Gd_{43}-\Gd_{45}
\Bigr]
\biggr]\non\\
&&
\phantom{\dot G(12)}
+\half \Gund_{45}
\biggl[
\Gd_{3k}
\Bigl[
\slG_{4i}-\slG_{5i}
\Bigr]
+\slG_{3i}
\Bigl[
\Gd_{34}-\Gd_{35}
\Bigr]
\biggr]
\Biggr\rbrace
+ 9 \,\,{\rm perm.}
\non\\
Q_5^{32}&=&
\Gd(123)\Gd(45) + 9 \,\,{\rm perm.}
\non\\
Q_5^{22}&=&
\Gd(12)\Gd(34)\Gd_{5i}
+ 14 \,\,{\rm perm.}
\label{5photoncoeff}
\ear\no
Again we have an over-counting here; $Q_5^{32}$ is
contained once in both $Q_5^3$ and $Q_5^2$, and
$Q_5^{22}$ is contained twice in $Q_5^2$.
And again every term appearing in this decomposition
is separately gauge invariant. Let us consider only the
least trivial case, which is a replacement of, say,
$\varepsilon_3$ by $k_3$ in 
(the un-permuted term of) $Q_5^2$. This leads to
the following total derivative,

\bear
&&
\partial_3
\Bigl[
\dot G(12)
\Gd_{4j}\Gd_{5k}
\expshort
\Bigr]
+\half
(\partial_3-\partial_4)
\Bigl[
\dot G(12)
\Gund_{34}
\Gd_{5k}
\expabb
\Bigr]
\nonumber\\&&
+\half
(\partial_3-\partial_5)
\Bigl[
\dot G(12)
\Gund_{35}
\Gd_{4k}
\expabb
\Bigr]
+\half
\partial_5
\Bigl[
\dot G(12)
\Gund_{34}
\bigl(
\Gd_{53}-\Gd_{54}
\bigr)
\expabb
\Bigr]
\nonumber\\&&
+\half
\partial_4
\Bigl[
\dot G(12)
\Gund_{35}
\bigl(
\Gd_{43}-\Gd_{45}
\bigr)
\expabb
\Bigr]
+\half
\partial_3
\Bigl[
\dot G(12)
\Gund_{45}
\bigl(
\slG_{4i}-\slG_{5i}
+\Gd_{34}-\Gd_{35}
\bigr)
\expabb
\Bigr]
\non\\
\label{Q5totderiv}
\ear\no

Comparing the 2 - and 3 - tails appearing in (\ref{5photoncoeff}) with
our results for $N=2,3$ we note that there is a simple relation between
$T_2,T_3$ and  $Q_2,Q_3$. The tail $T_i$  can be obtained from $Q_i$, in
its un-decomposed form, by rewriting $Q_i$ in the tail variables, and
then extending the range of all dummy indices 
to run over the complete set of variables
$\tau_1,\ldots,\tau_5$. It is not difficult to see that this relation
generalizes to an arbitrary  $Q_m, T_m$.  Consider (the unpermuted term
of) $Q_N^2$, which has a 2 - cycle $\dot G(12)$ and a tail
$T_{N-2}(3\ldots N)$. It suffices to consider those terms in $Q_N$
having a $\varepsilon_1\cdot k_2\varepsilon_2\cdot k_1$ as their
$Z_2(12)$ -- component. From the master formula
eq.(\ref{scalarqedmaster}) one infers that for this part of $Q_N^2$ the
partial integration procedure can have involved only  partial
integrations over the tail variables $\tau_3,\ldots,\tau_N$. Thus  the
calculation of $T_{N-2}$ and the lower order calculation of $Q_{N-2}$
are identical as far as the tail indices are concerned. The presence of
the cycle variables for the tail makes itself felt  only through an
extension of the momentum sums in the master formula, leading to the
stated extension rule for dummy indices. The same type of argument shows
that the structure of $T_m$ does not  depend on the number and length of
the cycles it multiplies.

At this point it should be noted that
every term in $Q_N$ must have 
at least one cycle factor (this is
a combinatorial consequence
of the fact that each such term contains
a total of $2N$ indices, of which only $N$
are different).
Thus the maximal tail occurring in $Q_N$
has length $N-2$. The above connection
between $T_N$ and $Q_N$ thus allows us to write down,
without going through the partial integration
procedure again, $Q_6$ as follows,

\be
Q_6
=
Q_6^6 + Q_6^5 + Q_6^4 + Q_6^3
+ Q_6^2
-Q_6^{42} 
- Q_6^{33} - Q_6^{32}
- Q_6^{22}
+ Q_6^{222}
\non\\
\label{6photon}
\ee\no
where
\bear
Q_6^6
 &=& 
\dot G(123456)
+  {\rm permutations} 
\quad\Bigl(
{5!\over 2} = 60 {\rm \,\,in\,\, total}
\Bigr)
\non\\
Q_6^5 &=&
\dot G(12345)
T_1(6)
+  {\rm perm.}
\quad\Bigl(
{4!\over2}{6\choose 1}
=72
 {\rm \,\,in\,\, total}
\Bigr)
\non\\
Q_6^4 &=&
\dot G(1234)
T_2(56)
+  {\rm perm.}
\quad\Bigl(
45
 {\rm \,\,in\,\, total}
\Bigr)
\non\\
Q_6^3 &=&
\dot G(123)
T_3(456)
+  {\rm perm.}
\quad\Bigl(
20
 {\rm \,\,in\,\, total}
\Bigr)
\non\\
Q_6^2 &=&
\dot G(12)
T_4(3456)
+  {\rm perm.}
\quad\Bigl(
15
 {\rm \,\,in\,\, total}
\Bigr)
\non\\
Q_6^{42}&=&
\Gd(1234)\Gd(56) 
+  {\rm perm.}
\quad\Bigl(
45
 {\rm \,\,in\,\, total}
\Bigr)
\non\\
Q_6^{33}&=&
\Gd(123)\Gd(456)
+  {\rm perm.}
\quad\Bigl(
10
 {\rm \,\,in\,\, total}
\Bigr)\non\\
Q_6^{32}&=&
\Gd(123)\Gd(45)T_1(6)
+  {\rm perm.}
\quad\Bigl(
60
 {\rm \,\,in\,\, total}
\Bigr)\non\\
Q_6^{22}&=&
\Gd(12)\Gd(34)T_2(56)
+  {\rm perm.}
\quad\Bigl(
45
 {\rm \,\,in\,\, total}
\Bigr)
\non\\
Q_6^{222}&=&
\Gd(12)\Gd(34)\Gd(56)
+  {\rm perm.}
\quad\Bigl(
15
 {\rm \,\,in\,\, total}
\Bigr)
\label{6photoncoeff}
\ear\no
Here the only new ingredient, $T_4$, according to the
above is related to the un-decomposed $Q_4$ of
eq.(\ref{Q4}) simply by a relabelling, and an
extension of the range of
all dummy indices to run from $1$ to $6$. 

Note that the integrand is
not yet quite suitable for the
application of the cycle substitution
rules, since the tails still contain cycles. 
For this purpose, one should further rewrite 
$Q_6$ as

\be
Q_6=
\hat Q_6^6 + \hat Q_6^5 + \hat Q_6^4 + \hat Q_6^3
+ \hat Q_6^2
+\hat Q_6^{42} 
+ \hat Q_6^{33} + \hat Q_6^{32}
+ \hat Q_6^{22}
+ \hat Q_6^{222}
\label{Q6altern}
\ee\no
where the ``hat'' on a term means that
the range of the dummy indices appearing in its tail
has been restricted so as to eliminate all
additional cycles. This also removes the over-counting,
so that now all coefficients are unity.

It is now obvious that in this way one arrives
at a canonical permutation symmetric
version of the Bern-Kosower integrand for
the one-loop $N$ - photon/gluon amplitude. 
Moreover, this integrand naturally
decomposes into gauge invariant partial
amplitudes. To see the gauge invariance, note that 
at every step of the recursion only one new structure
appears in $Q_N$, namely $T_{N-2}$. 
The
separate gauge invariance of all terms except
$Q_N^2$ can be inferred from the gauge invariance
of lower order terms, since the total length  
of the cycles multiplying a given tail is
clearly not relevant for this analysis.
Since the complete integrand must be gauge
invariant so must be $Q_N^2$.

\noindent
In the abelian case, 
the final parameter integral gives the
complete $N$ - photon amplitude,

\bear
\Gamma
[k_1,\varepsilon_1;\ldots ;k_N,\varepsilon_N]
&=&
{(-g)}^N
{(2\pi )}^D\delta (\sum k_i)
{\dps\int_{0}^{\infty}}{dT\over T}
{[4\pi T]}^{-{D\over 2}}
e^{-m^2T}
\non\\
&&\times
\prod_{i=1}^N \int_0^T 
d\tau_i\,
Q_N(\dot G_{Bij})
\exp\biggl\lbrace\sum_{i<j=1}^N 
G_{Bij} k_i\cdot k_j
\biggr\rbrace
\label{Nphotonamplitude}
\ear\no
with no need to add permuted terms.

In the non-Abelian case the integration region is
restricted by the colour ordering as in
(\ref{scalarqedmaster}), so that one must explicitly
sum over all non-cyclic permutations of the
$N$ gluons.
Moreover,
additional
boundary contributions are generated in the
partial integration process. Those correspond
to the ``tree part'' of the Bern - Kosower rules,
and in the effective action picture merely contribute to the
covariantization of the main term ~\cite{strassler2}.
Their structure will be discussed elsewhere.

We expect the above construction of the
Bern - Kosower integrand to be useful in
future applications of the Bern-Kosower formalism
beyond the five - gluon amplitude.
Another possible application is the calculation
of the QED four - photon amplitude with all four
legs off-shell, which has not yet been done 
to the knowledge of the author. The gauge
structure of this amplitude was analyzed in
~\cite{cotopi}, however the decomposition into gauge
invariant 
partial amplitudes  given there is not identical
with the one proposed here.
Note that the partial integration procedure makes the
finiteness of this amplitude manifest at the integrand level,
since in contrast to $P_4$ all terms in $Q_4$ have already
four external momenta factored out.

Moreover our considerations
should also have consequences at the multiloop level,
since eqs.(\ref{scalarqedmaster}),(\ref{Nphotonamplitude})
are valid off-shell
\footnote{
The off-shellness
was not obvious in the original derivation of the
Bern - Kosower 
master formula, since for the initial string amplitudes
the requirement of conformal invariance forces the external states to
be on-shell.},
and thus can serve as a starting point
for the construction of multiloop amplitudes 
~\cite{ss23}. In terms
of Feynman diagrams, the
partial integration procedure then effectively
induces an intricate re-shuffling of terms
between numbers of diagrams of various
topologies, destroying the initial
connection to standard
Feynman parameter integrals.
The properties of the canonical integrand at the 
multiloop level are presently under investigation.

\noindent
{\bf Acknowledgements:}
The author would like to thank Z. Bern and M. J. Strassler for
various informations, and D. Fliegner for help with computer work.
The present investigation made use of the symbolic computation
program M ~\cite{M}.

\vfill\eject

\end{document}